\documentclass[amsmath,amssymb,amsbsy,prl,nofootinbib,twocolumn,tightenlines,superscriptaddress,floatfix]{revtex4-2}
\usepackage{amsfonts} 
\usepackage{amsmath} 
\usepackage{amssymb}
\usepackage{blindtext}
\usepackage{lineno}
\usepackage{cancel}
\usepackage{bm}
\usepackage[usenames, dvipsnames]{color} 
\usepackage{dcolumn}
\usepackage{epic} 
\usepackage{epsfig}
\usepackage{eucal} 
\usepackage{esdiff}
\usepackage{graphicx}
\usepackage[breaklinks,colorlinks = true,linkcolor = blue,urlcolor=blue,citecolor=red]{hyperref}

\newcommand{\bx}{\boldsymbol{x}}
\newcommand{\bX}{\boldsymbol{X}}
\newcommand{\bu}{\boldsymbol{u}}
\newcommand{\bV}{\boldsymbol{V}}
\newcommand{\br}{\boldsymbol{r}}

\usepackage{hyperref}
\usepackage{array}    
\usepackage{float}

\newcommand{\icts}{International Centre for  Theoretical Sciences, Tata Institute of Fundamental Research,  Bangalore 560089, India}

\begin{document}
\title{The significance of two-way coupling in two-dimensional, dusty turbulence}
\author{Harshit Joshi$^\clubsuit$}
\email{hj921999@gmail.com}
\affiliation{\icts}
\author{Amal Manoharan$^\clubsuit$}
\email{amalmanoharan1994@gmail.com}
\affiliation{\icts}
\author{Samriddhi Sankar Ray}
\email{samriddhisankarray@gmail.com}
\affiliation{\icts}

\begin{abstract}
The significance of small-scale forcing of particles on the carrier 
two-dimensional turbulent flow has been shown to influence the spectral 
scaling properties of the carrier fluid. We investigate possible consequences 
of such two-way coupling in a turbulent suspension of inertial particles 
through one- and two-point Eulerian and Lagrangian statistics. In particular, 
we find signatures of enhanced intermittency in the vorticity distributions. 
We characterize the changes in the small-scale geometry of the flow via the 
Okubo--Weiss parameter. Finally, we examine the scaling properties of the 
second-order vorticity structure functions and find a non-trivial form of 
scale-invariance at finite mass loading. 
\textcolor{black}{Motivated by these observations, we propose an effective 
multiscale forcing framework in which particle feedback is modeled as a 
spatially localized small-scale forcing. This dual-scale forcing captures the 
emergence of modified spectral scaling and provides a minimal Eulerian 
description of particle-laden turbulence that reproduces key statistical 
signatures of the system.}
\end{abstract}

\maketitle

\def\thefootnote{$\clubsuit$}\footnotetext{These authors contributed equally to this work}\def\thefootnote{\arabic{footnote}}

Turbulent transport --- the dynamics of particulate matter in turbulence --- is
ubiquitous~\cite{VothReview,BecReview}. In particular, the understanding of
settling~\cite{WangSettling1993,BecGravSettling2014,AnandSettling2020},
collision~\cite{Sundaram1997,AyalaPart12008,Saw2014,James2017,PicardoCollisions2019},
coalescence and aggregation~\cite{Kostinski2005,BecCoalescence2016} in the
context of pollutant dispersion, embryonic
proto-planets~\cite{Wetherill1989,Lissauer1993}, and the microphysics of warm
and cold clouds~\cite{Pinsky1997,Falkovich2002,Shaw2003} is particularly
important. The latter assumes special significance, as these processes form building blocks
of large-scale climate models, which still under-resolve dynamics at cloud scales~\cite{Liu2023,Perkins2024}. 
The thrust of many studies, including the present one, has been within the ambit of small enough particles: particle
sizes of length $a$, such that $a$ is the radius of spherical droplets or the length of the semi-major axis of anisotropic spheroids, are much smaller than
the relevant small scales $\eta$ of turbulence~\cite{VothReview,BecReview}.
While this assumption is somewhat restrictive --- indeed, an equally large body
of work exists for particles larger than the $\eta$ --- it is found to be
generally valid for several problems, including those that relate to the
microphysics of clouds~\cite{Bodenschatz2010,GrabowskiReview,Ravichandran2020}.  For example,
the typical radius of water droplets or the semi-major axis of spheroidal ice
crystals $10 \mu m \lesssim a \lesssim 100 \mu m$ lie well below the
Kolmogorov scale $\eta \approx 1 mm$ in a typical, turbulent atmospheric
setting. 

One of the achievements of the large body of work on such sub-Kolmogorov 
particles, with an eye on the cloud microphysics problem, has been to underline 
the enormity of the role played by a key fingerprint of turbulence --- intermittency --- 
in deciding the eventual outcome of processes such as coalescence and aggregation~\cite{BecCoalescence2016}. 
An example of this is the direct calculation of the coalescence kernel, within 
the framework of the Smoluchowski equation, where the intermittency exponent shows up 
as a dominant term~\cite{BecCoalescence2016}.

The results discussed suggest that the taming or tinkering of fluid intermittency could result in some revision of the conclusions drawn above. In
particular, such studies assume, quite legitimately, that dilute suspensions of
$a \ll \eta$ particles imply that particles have no effect on the carrier flow:
the fluid-particle interactions are one-way coupled, with only the flow
affecting the particles and not vice versa. It is tempting to consider whether this restriction is relaxed~\cite{Ferrante2003,Eaton2009,BecDusty2017,GualtieriDusty2017,Nath2025} and the particles also feed the flow. How might (a) the properties of the
turbulent flow change, and (b) how do the dynamics of the particles themselves
alter in this two-way coupled regime?

This question was addressed, in part, in a paper by Pandey, Perlekar,
and Mitra~\cite{PandeyDusty2019}. They showed, through state-of-the-art Direct Numerical Simulations
(DNSs) of Stokesian particles suspended in a two-dimensional (2D) turbulent
flow, that the effect of the two-way coupling leads to a distinct change in the
kinetic energy spectrum of the fluid with an emergent scaling regime associated
with a scaling exponent, which depends on the Stokes number $St$ and the mass
loading $\phi_m$, and hence non-universal. Conversely, the two-way coupling
also results in significant changes in the nature of particle clustering: the
commonly accepted measure of clustering, the correlation dimension $D_2$
measured via the probability $P^<(r) \sim r^{D_2}$~\cite{BecD22003,BecD22005} of two particles being
within a distance $r$, is shown to be sensitive to $\phi_m$ and converges, as
one would expect, to the one-way coupled, dilute regime values, as $\phi_m \to
0$.

This work~\cite{PandeyDusty2019} opens up several interesting questions
which relate to both the fluid and the particles. One such question, which we
partially address in this paper, concerns the strength of the
small-scale forcing of the particulate phase on the carrier flow,
the turbulent flow itself. This is especially important given our current understanding 
of the dominant role of turbulence intermittency 
in processes such as accelerated aggregation.

Our work is based on the dusty turbulence model 
of Pandey \textit{et al.}~\cite{PandeyDusty2019} where heavy particles 
are immersed in a two-dimensional (2D) turbulent flow. Given that the particles are 
smaller than the small length scales of the flow, the force exerted by the fluid on the
$i^{\rm th}$ particle is best approximated by the linear Stokes drag. Thus, its 
dynamics --- in terms of its position $\bX$ and velocity $\bV$ --- are described by

\begin{equation}
\label{eq:Xdot}
\frac{d\bX_i}{dt} = \bV_i(t), \quad
\frac{d\bV_i}{dt} = \frac{1}{\tau_p}[\bu(\bX_i, t) - \bV_i],
\end{equation}

where $\tau_p \textcolor{black}{ = \frac{2\rho_p a^2}{9\rho_f \nu}}$ is the Stokesian relaxation time scale of the individual 
particle. These particles are immersed in a turbulent flow, with kinematic 
viscosity $\nu$ and the coefficient of Ekman friction $\alpha$, whose velocity field $\bu(\bX, t)$ obeys 
the 2D incompressible Navier--Stokes equations.
In two-dimensional systems, it is most convenient to write the 
Navier--Stokes equations in terms of the (pseudo)-scalar vorticity field $\omega =
\boldsymbol\nabla \times \bu$:

\begin{multline}
\label{eq:wEqn}
\partial_t\omega(\bx, t) + \boldsymbol{u}\cdot \boldsymbol\nabla \omega(\bx, t) = \nu \nabla^2 \omega(\bx, t)   -\alpha\, \omega(\bx) \\
+  f(\bx) 
+ \boldsymbol\nabla\times\boldsymbol{F}^{d}(\bx, t).
\end{multline}

The last term $\boldsymbol\nabla\times\boldsymbol{F}^{d}(\bx, t)$ is the critical, additional forcing term 
arising from the effect of the particles on the fluid, ensuring the two-way coupling of the model 
Eqs.~\eqref{eq:Xdot}-\eqref{eq:wEqn} for dusty turbulence. The form of this additional forcing on the fluid phase 
is readily deduced by demanding that the fluid-particle system be momentum-conserving:
\begin{equation}
\label{eq:Fd}
\boldsymbol{F}^d(\bx, t) = \sum_{i=1}^{N_p} \frac{m}{\tau_p \rho_f} [\bV_i(t) - \bu(\bX_i, t)] \delta^2(\bx - \bX_i),
\end{equation}
where $m$ is the mass of each of the total number $N_p$ of particles. 
The relevant dimensionless numbers, apart from the Reynolds number $Re$, are the Stokes number $St$ and the mass loading parameter $\phi_m = m N_p/(\rho_f L^2)$, where $\rho_f$ is the fluid density and $L$ the 
domain length.

Besides the particle feedback force on the fluid, we also consider an external,
large-scale forcing $f(\bx)$ to drive the system into a non-equilibrium statistically steady state, even in the 
absence of the particles.

\begin{figure}
\includegraphics[width = 1\linewidth]{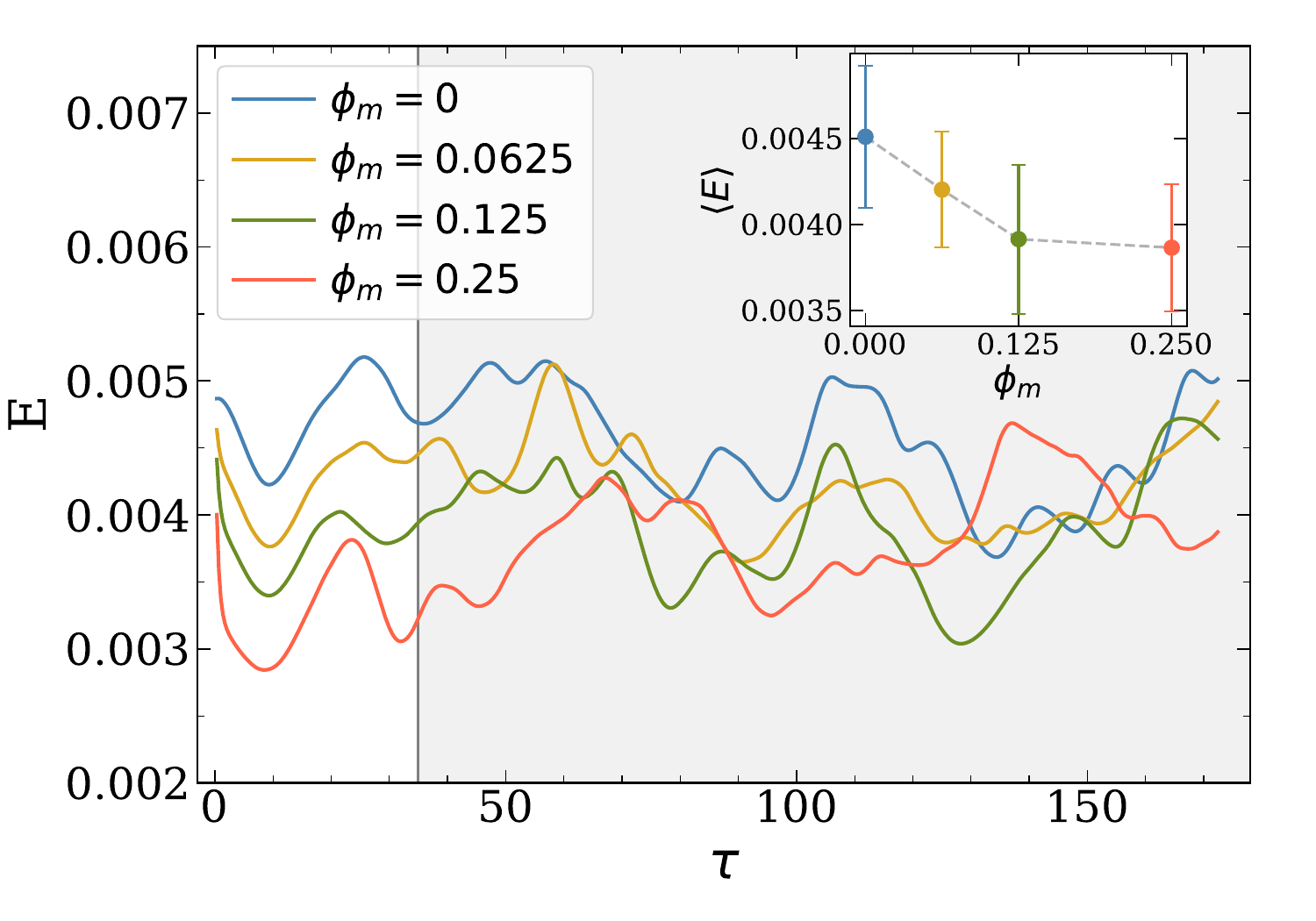}
\caption{Fluid kinetic energy $E$ versus non-dimensional time $t/\tau_\eta$ for different mass loadings $\phi_m$. The shaded region indicates the statistically steady state. 
\textcolor{black}{Inset: Mean kinetic energy $\langle E \rangle$ with standard deviation (error bars) in the steady state as a function of $\phi_m$.}}
\label{fig:energyTime}
\end{figure}

We solve the dusty turbulent flow through pseudo-spectral, fully de-aliased
direct numerical simulations (DNSs) on a doubly-periodic $L^2 = 4\pi^2$ domain
with a second-order Runge-Kutta scheme to evolve both fluid and particles in
time~\cite{PerlekarPersistence2011}. We use $1024^2$ collocation points to discretize the simulation
domain and choose $\nu = 10^{-5}$ and $\alpha = 0.01$. The external forcing
$f(\bx) = -f_0k_f \cos(k_f y)$ acts at wavenumber $k_f = 4$ with an amplitude
$f_0 = 5 \times 10^{-3}$; the initial condition for the vorticity field is chosen 
to be $\omega(\bx, 0) = -f_0k_f\nu[\cos(k_f x)+\cos(k_f y)]$. 
\textcolor{black}{We perform additional simulations with $N^2 = 2048^2$ for packing fractions where the 
effect of particles is strongest, to study the effect of grid resolution 
on the particle feedback force. The main panels and figures that follow 
are from the $N^2 = 1024^2$ simulations, and, where indicated, the insets show results from the $N^2 = 2048^2$ 
simulations for comparison.}

Along with the fluid phase, we immerse $N_p$ particles at random locations and with zero velocities, with $N_p$ in the range $6.25\times 10^4 \leq N_p \leq 5\times10^{5}$, resulting in a mass loading $0.0325 \leq \phi_m \leq 0.25$. We also use different families of particles with 
different Stokes numbers $St = \tau_p/\tau_\eta$ in the range $0.17 \leq St \leq 1.67$ by varying $\tau_p$, where the small 
fluid time scale $\tau_\eta = \sqrt{\nu/\epsilon}$ and $\epsilon$ is the mean energy dissipation rate. In what follows, 
we present results for $St = 0.67$ (and sometimes for Lagrangian particles with $St = 0$), since, as seen 
before~\cite{PandeyDusty2019}, such Stokes numbers result in the strongest effects of particle feedback 
on the flow. For Stokes numbers much smaller or larger, the effects we report persist but weaken.

We now come to the feedback force of the particles on the fluid. 
We follow the prescription of earlier studies~\cite{PeskinImmersed2002, PandeyDusty2019} 
to construct the two-dimensional $\delta^2$ function on grids of linear
dimension $h$ from a one-dimensional $\delta$ function

\begin{equation}
\label{eq:deltaNum}
\delta(x) = 
\begin{cases}
    \dfrac{1}{4h}\left\{1+\cos\left[\dfrac{\pi x}{2h}\right] \right\}, & |x|\leq 2h, \\
    0, & \text{otherwise}.
\end{cases}
\end{equation}
With this prescription, the fluid velocity at the particle position $\bX$ is interpolated as
\begin{equation}
\label{eq:uAtX}
\bu(\bX, t) = \sum_{\bx} \bu(\bx, t) \delta^2(\bx-\bX) h^2.
\end{equation}
{\color{black}where 
\begin{equation}
	\delta^{2}(\mathbf{x} - \mathbf{X}) = \delta(x - X)\delta(y - Y).
\end{equation}
}
The localized cosine discrete $\delta$-function defined in equation \eqref{eq:deltaNum} is used both for spreading particle forces to the grid and
for interpolating the fluid velocity to particle positions, as in equation \eqref{eq:uAtX}, following the standard immersed boundary method formulation
\cite{PeskinImmersed2002}.

\begin{figure*}
	\includegraphics[width=0.61\linewidth]{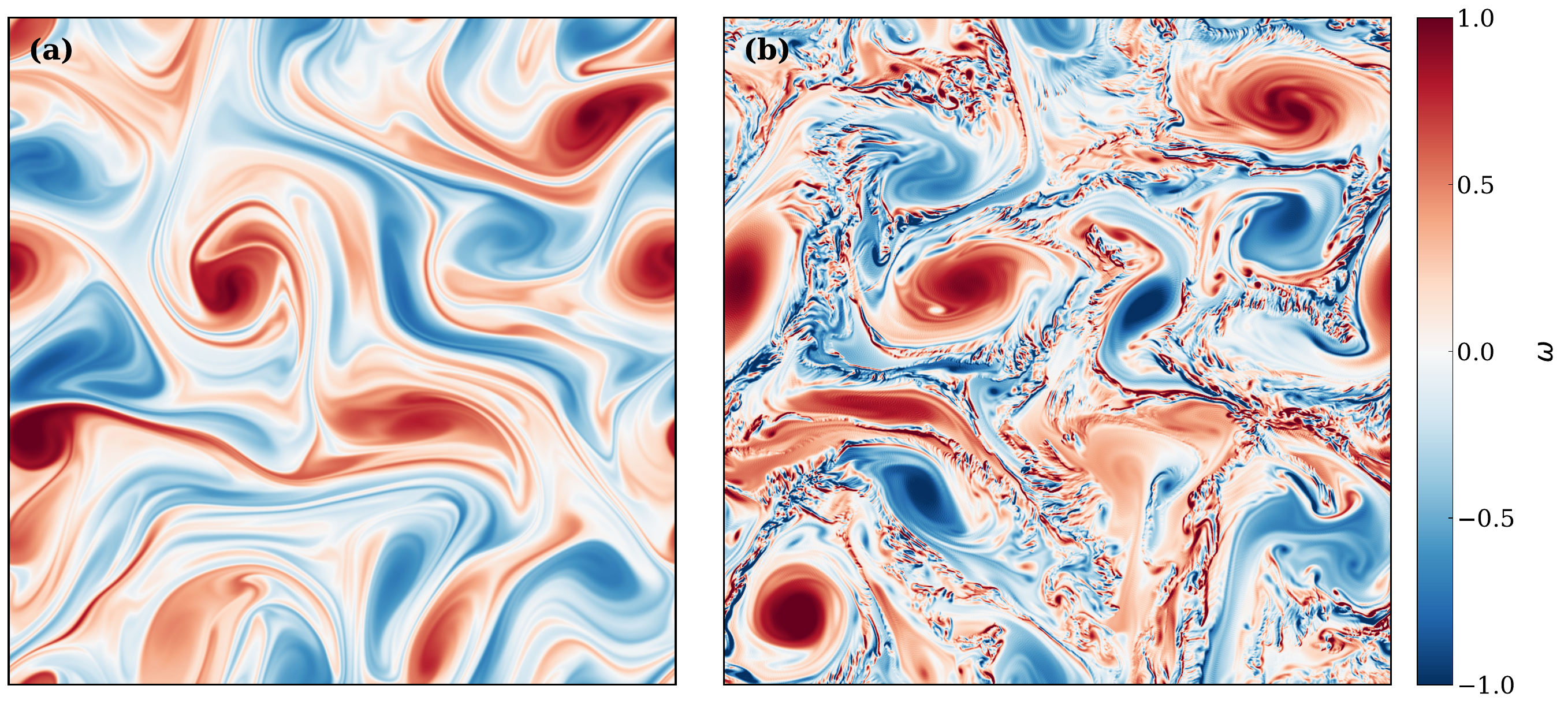}
	\includegraphics[width=0.36\linewidth,height=4.8cm]{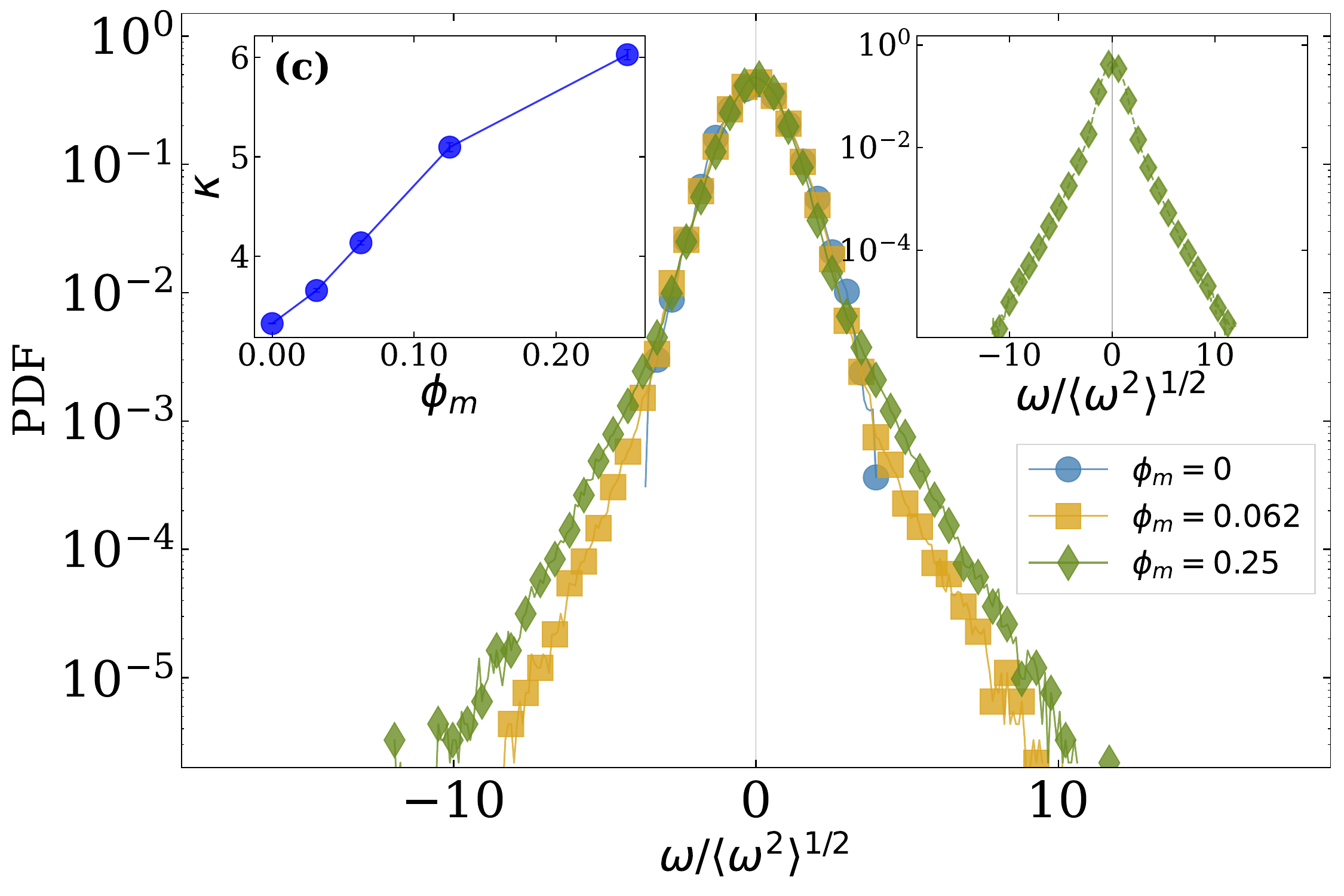}
\caption{Pseudo-color plots of the vorticity field for (a) $\phi_m = 0$ and (b) $\phi_m = 0.125$, with $N_p = 2.5 \times 10^5$ particles at $St = 0.67$. 
(c) Probability density functions (PDFs) of $\omega(x)$, normalized by their variance, for different $\phi_m$; the dashed black curve indicates a Gaussian. 
(Left inset) Kurtosis $\kappa$ as a function of $\phi_m$ (error bars via bootstrapping). 
\textcolor{black}{(Right inset) PDF for $\phi_m = 0.25$ from a $N^2 = 2048^2$ simulation, showing the same qualitative behavior as the $N^2 = 1024^2$ results in the main panel.}}
\label{fig:Eu1ptVort}
\end{figure*}

We begin simulations without particles to first achieve a statistically steady state of a
two-dimensional turbulent flow. We then immerse particles in this flow. 
The time evolution of this combined system results in a slightly different (for $\phi_m \neq 0$) non-equilibrium,
statistically steady state, with a typical Taylor-scale based Reynolds number
$Re \approx 1866$. In Fig.~\ref{fig:energyTime} we show a plot of the evolution
of the total kinetic energy $E$ vs the non-dimensional time $\tau =
t/\tau_\eta$ for different mass-loading after particle immersion. The shaded region indicates the
stationary regime, which is achieved rather quickly; for $\phi_m = 0$ the steady state of the turbulent fluid phase is, of course, unchanged. 
\textcolor{black}{We estimate the mean kinetic energy as the time average of the 
measured kinetic energy $E$ in the steady state, as indicated by the shaded region in 
Fig.~\ref{fig:energyTime}, and the fluctuation as the standard deviation of $E$ over this time window. 
We show the variation of this mean kinetic energy $\langle E \rangle$, along with its error bars, 
as a function of $\phi_m$ in the inset of Fig.~\ref{fig:energyTime}.}

How do the small-scale forces, coming from the immersed particles, affect the
structure of the carrier flow? In Fig.~\ref{fig:Eu1ptVort}, we compare the fields
for (a) the particle feedback force switched off ($\phi_m  = 0$) and (b) the
case where the particle feedback is included for a mass loading $\phi_m = 0.25$
and $St = 0.67$. Visually, as observed earlier~\cite{PandeyDusty2019}, the effect of particles on the flow leads to the conspicuous development of small-scale structures, best seen by comparing pseudo-color plots of the vorticity field with and
without particle feedback.

This generation of more intense small-scale structures is best captured in the
probability distribution function (PDF) of the vorticity field with and without
a finite mass-loading. We alert the reader that all simulations are performed
with particles --- the case of no mass-loading ($\phi_m = 0$) corresponds to simulations where the particles are passive and feedback on the fluid is turned off: $\boldsymbol{F}^d(\bx, t) = 0$ in the Navier-Stokes equation.

In Fig.~\ref{fig:Eu1ptVort}(c), we show representative semilog plots of the PDFs
for the (normalised) vorticity fields for \textcolor{black}{$St = 0.67$ at different values of $\phi_m$, as well as contrast this with a Gaussian distribution indicated by a black dashed line.} The visual cue stemming from panels (a) and (b) shows up in
the strong deviation of these PDFs for $\phi_m \neq 0$ from the nearly-Gaussian
$\phi_m = 0$ case. Indeed, the wide tails, which seem to grow wider with
increasing $\phi_m$, are strongly reminiscent of what one would expect for a
strongly intermittent system, albeit, surprisingly, at the level of
single-point statistics. We quantify this departure from a nearly Gaussian
field through the kurtosis $\kappa$ as a function of $\phi_m$
(Fig.~\ref{fig:Eu1ptVort}(c), left inset). The kurtosis indeed shows a monotonic
increase with increasing mass-loading, with $\kappa \gtrsim 3$ as $\phi_m \to
0$ and $\kappa \gtrsim 6$ for our largest $\phi_m$.

\textcolor{black}{A key numerical issue concerns the sensitivity of the
vorticity PDF to grid resolution, given that the particle feedback force
involves a discrete approximation of the Dirac delta function. To address this,
we perform additional simulations at a higher resolution of $N^2 = 2048^2$.}

\textcolor{black}{We consider the physically
relevant limit in which the delta function is approximated over a fixed number
of grid points, so that its physical support shrinks with increasing
resolution, consistent with the modeling of point-particle forcing. In this
case, exact overlap of the PDFs across resolutions is not expected, since the
forcing becomes increasingly localized and its small-scale content changes with
resolution. However, robustness requires that the functional form of the
distributions be preserved.}

\textcolor{black}{In the right inset of Fig.~\ref{fig:Eu1ptVort}(c), we show a
semilog plot of the vorticity PDF for the extreme case $\phi_m = 0.25$ and $St
= 0.67$ from the $2048^2$ simulations. The distributions exhibit the same tail
behavior and comparable deviations from Gaussianity as in the lower-resolution
runs. This indicates that the observed intermittency and non-Gaussian features
are not artifacts of the grid-scale regularization of the delta function, but
are robust with respect to grid refinement. We, however, expect weak sensitivity
to the precise kernel form provided it remains localized; a systematic
comparison of different kernel shapes is beyond the scope of the present
study.}

We now explore the question of intermittency through the more conventional approach 
of two-point correlations via the $p$-th order structure functions $S_p^\omega(r) \equiv \langle |\delta_r\omega|^p \rangle = \langle |\omega(\bx+\br) - \omega(\bx)|^p \rangle \sim r^{\zeta_p}$, 
where the average $\langle \cdot \cdot \cdot \rangle$ is defined over the spatial positions 
$\bx$ and all directions $\br/|\br|$ for a given $r=|\br|$ in the inertial range.

\begin{figure*}
	\includegraphics[width=1\linewidth]{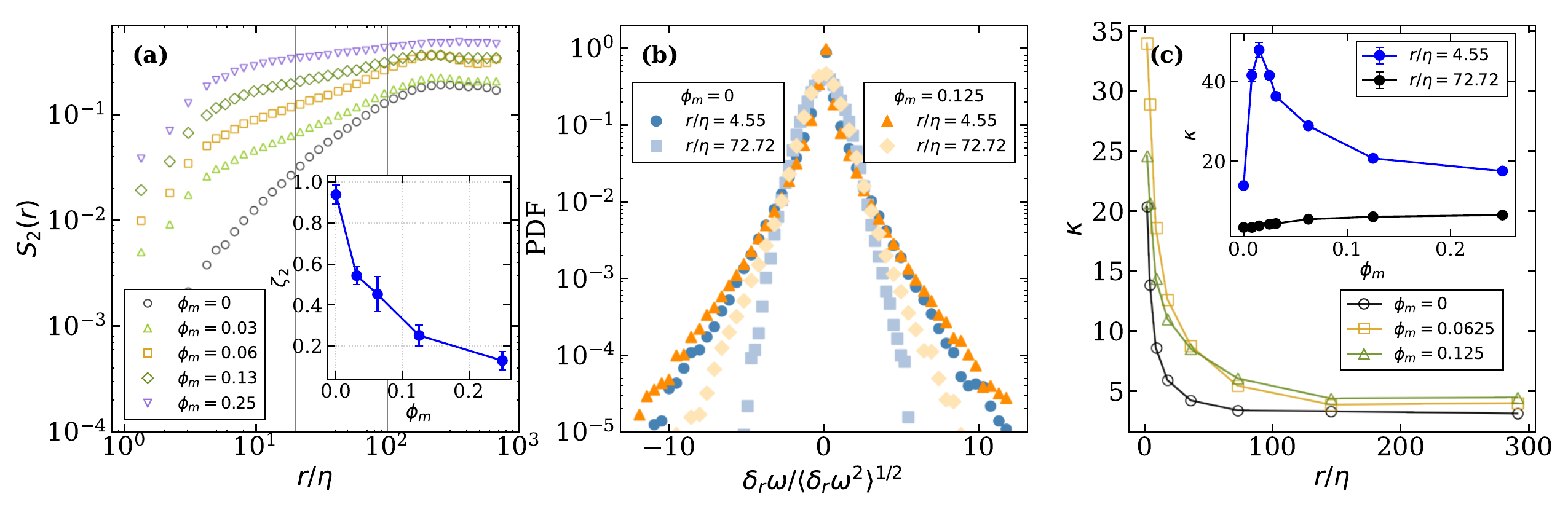}
	\caption{(a) Second-order structure function $S_2(r)$ versus $r/\eta$ for different $\phi_m$; vertical lines indicate the dominant scaling range $S_2 \sim r^{\zeta_2}$. 
(Inset) $\zeta_2$ as a function of $\phi_m$. 
(b) PDFs of vorticity increments for two values of $\phi_m$ and separations. 
(c) Scale-dependent kurtosis $\kappa$ for different $\phi_m$. 
(Inset) $\kappa$ at fixed separations $r^*/\eta = 4.55$ and $r^*/\eta = 72.72$ as a function of $\phi_m$.}
\label{fig:intermittency}
\end{figure*}

In Fig.~\ref{fig:intermittency}(a), we show a representative plot of second-order structure
functions for different mass-loading $\phi_m$ and $St = 0.67$. Two things
stand out. First, the scaling exponent seems to strongly depend on $\phi_m$
and, for large $\phi_m$, shows a saturation with $S_2 \sim r^0$. Second, it's not
obvious from such plots if there is a single scaling exponent or multiple scaling
exponents (in the forward cascade regime) because of the non-zero particle
feedback. We return to this point towards the end of this paper.

Nevertheless, there is \textit{a} dominant scaling regime as denoted by the pair
of vertical lines, from which we measure $\zeta_2$; the values of $\zeta_2$ as a function
of $\phi_m$ are shown in the inset of Fig.~\ref{fig:intermittency}(a). Curiously, our measurements suggest that as
$\phi_m$ becomes large, $\zeta_2 \gtrsim 0$, signaling a progressive loss of spatial
correlation in the vorticity field. This is not surprising given the evidence
already seen in Fig.~\ref{fig:Eu1ptVort}(b) and reported earlier.

\textcolor{black}{This loss of spatial correlations, as we discuss in greater
detail later, is perhaps a consequence of the injection of small-scale forcing
arising from particle feedback on the flow. In this sense, the resulting
decorrelation shares certain qualitative features with flows driven by
small-scale stochastic forcing, in that both lead to a progressive loss of
spatial coherence at small separations. However, it is important to emphasize
that, in the present case, the mechanism is entirely deterministic and stems
from the localized feedback forces exerted by inertial particles.}

But just how intermittent is the flow with the addition of the particle feedback? In order to measure this, we compute the PDFs of the vorticity increments, as
shown in Fig.~\ref{fig:intermittency}(b), for two different $\phi_m$ and two
different separations (normalised by $\eta$). Clearly, the effect of $\phi_m$
leads to a widening of the tails of such distributions, though not
dramatically, signaling the onset of stronger intermittency. We quantify this
in Fig.~\ref{fig:intermittency}(c) through plots of the scale-dependent kurtosis
$\kappa$ for different $\phi_m$. In the case of no particle feedback, the
kurtosis, at large separations, comes close to 3 --- indicating Gaussian
distributions --- whereas for $\phi_m \neq 0$, we observe $\kappa > 3$, with values
that increase with $\phi_m$. To quantify how $\kappa$ changes with
$\phi_m$ at a fixed $r/\eta$, in the inset, for two choices of separations, we
show how $\kappa$ varies with $\phi_m$. At small separations, there is a
distinct non-monotonic behaviour, with the largest $\kappa$ observed for small
$\phi_m$ and, with increasing $\phi_m$, a gradual convergence to the $\phi_m =
0$ values. However, for slightly larger $r/\eta$, this non-monotonicity vanishes
and $\kappa$ increases with $\phi_m$, with signs of an eventual saturation at
large $\phi_m$.

What does all of this mean for the dynamics of inertial particles? For Stokesian particles, the preferential sampling of the flow is strongly linked
to the local flow topology. Hence, if the feedback of the particles on the flow
($\phi_m \neq 0$) strongly influences the flow topology, it would have important
consequences for questions related to collisions and coalescences. A useful
measure of this local flow topology is the determinant of the velocity gradient
tensor: the Okubo-Weiss parameter $\Lambda = \det(\partial_j u_i)$. The sign of
$\Lambda$ provides a useful diagnostic to contrast the strain-dominated,
hyperbolic regions ($\Lambda < 0$) of the flow with those that are
vorticity-dominated, elliptic with closed streamlines ($\Lambda > 0$).

\begin{figure}
	\includegraphics[width=1.0\linewidth]{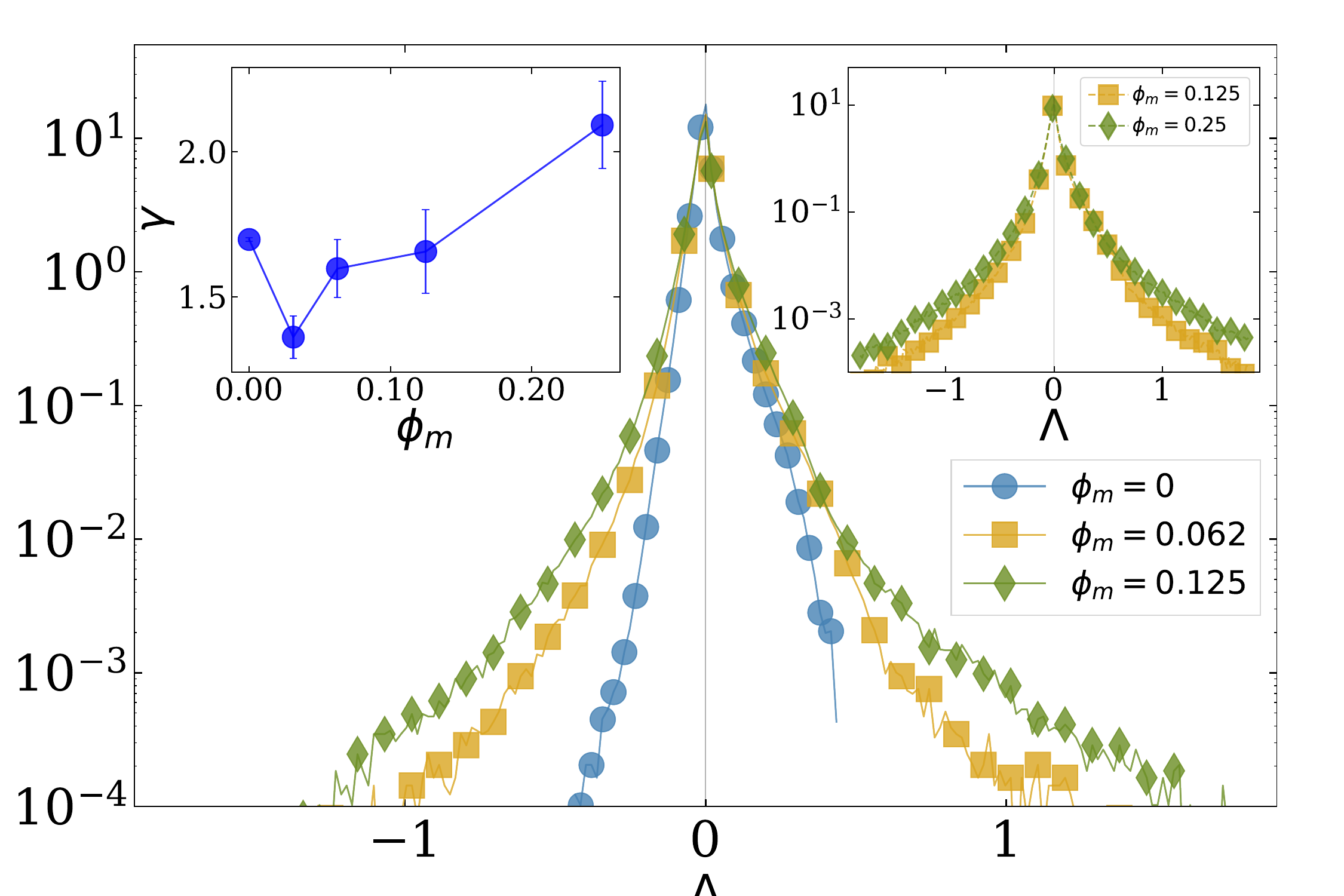}
\caption{PDFs of the Okubo--Weiss parameter $\Lambda$, computed along Lagrangian (tracer, $St = 0$) trajectories for different $\phi_m$ with $St = 0.67$. 
\textcolor{black}{(Right inset) Corresponding results from $N^2 = 2048^2$ simulations.}
(Left inset) Skewness $\gamma$ versus $\phi_m$ (error bars via bootstrapping).}
\label{fig:OWpdfSt0}
\end{figure}

We consider the suspension with $St = 0.67$ particles at different mass-loading $\phi_m$ and measure $\Lambda$ along 
particle trajectories for both Lagrangian (tracer) particles with $St = 0$ and heavy inertial particles with $St = 0.67$. 
In Fig.~\ref{fig:OWpdfSt0} we plot the PDF of $\Lambda$, as seen by Lagrangian particles, at different values of 
$\phi_m$, \textcolor{black}{with similar PDFs from a $N^2 = 2048^2$ simulation shown in the right inset}. 
This confirms the observation, already apparent in Fig.~\ref{fig:Eu1ptVort}(b), 
that the flow develops intense vortical and strain-dominated regions in local patches. Since these tracer particles 
sample the flow \textit{uniformly} and are a surrogate measure of the Eulerian statistics, 
the distribution maintains its positive skewness as vortical regions dominate two-dimensional turbulence. 
In fact, we find a marginal increase in the value of skewness with $\phi_m$ (Fig.~\ref{fig:OWpdfSt0}, left inset), 
although, within error-bars, it is not clear if this increase is significant.

\begin{figure}
	\includegraphics[width=1.0\linewidth]{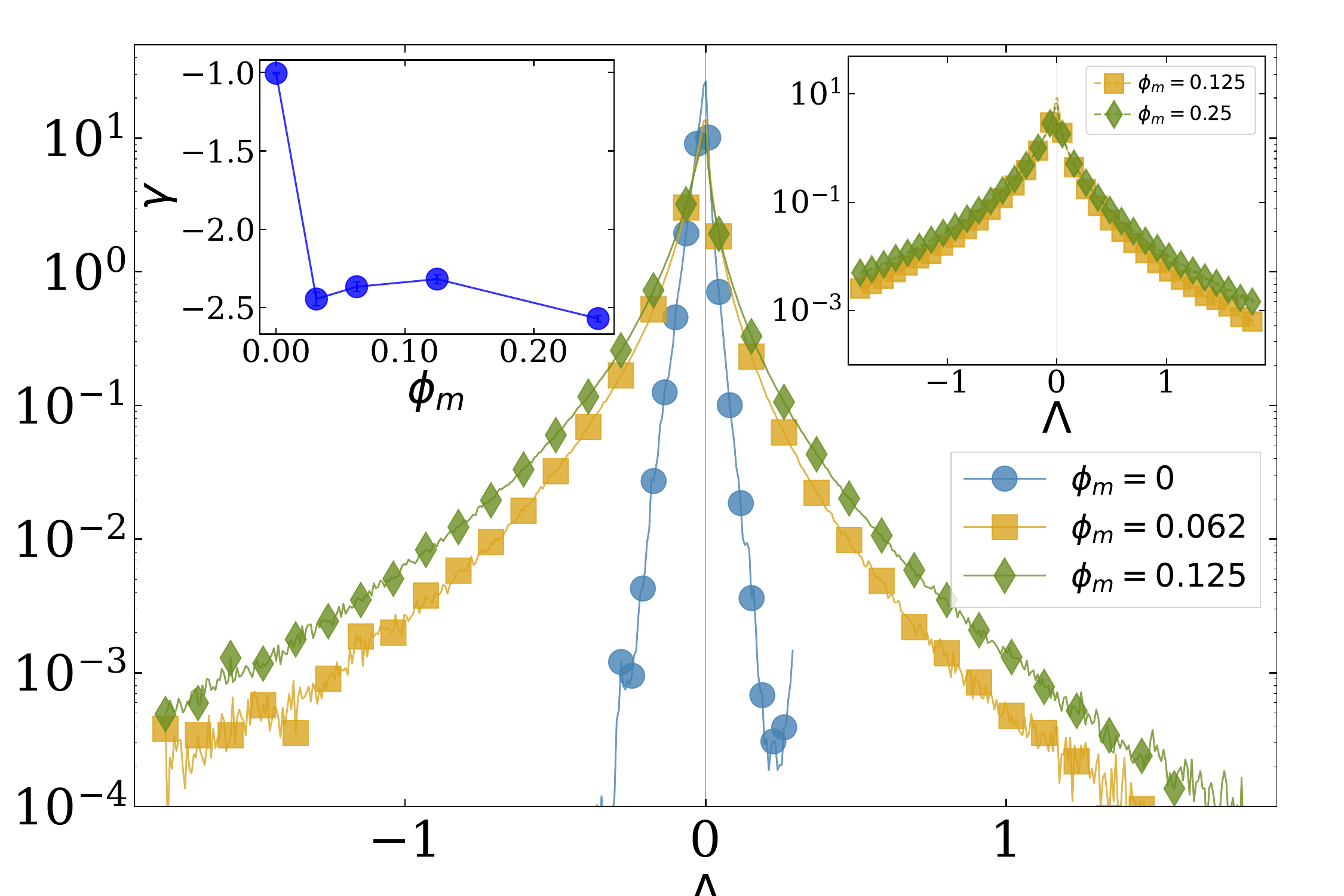}
\caption{PDFs of the Okubo--Weiss parameter $\Lambda$, computed along inertial particle trajectories ($St = 0.67$) for different $\phi_m$. 
Results from $N^2 = 1024^2$ (main panel) and 
\textcolor{black}{$N^2 = 2048^2$ (right inset).}
(Left inset) Skewness $\gamma$ with error bars.
}
\label{fig:OWpdfSt0p67}
\end{figure}

Inertial particles, on the other hand, sample the flow preferentially and tend to cluster in strain-dominated regions, resulting in negatively-skewed distributions. In Fig.~\ref{fig:OWpdfSt0p67} we calculate the PDFs of $\Lambda$ for 
$St = 0.67$ particles and for various $\phi_m$. While in the passive-particle limit, the skewness is negative, for 
$\phi_m \neq 0$, it becomes more pronounced (Fig.~\ref{fig:OWpdfSt0p67}, left inset) and saturates quickly with little or 
no variation in its value with changing $\phi_m$. \textcolor{black}{For comparison, in the right inset we show 
the distributions from the more resolved $N^2 = 2048^2$ simulations to underline the fact that this behaviour is not 
sensitive to the grid resolutions used.}

Our results are interesting for several reasons.  \textcolor{black}{First, the
Okubo--Weiss parameter measurements suggest that the addition of particle
feedback forces alters the small-scale statistics of the flow substantially.
This, in turn, may lead to a revision in the modelling of collisions and
coalescences when such feedback forces are not accounted for~\cite{Sundaram1997,AyalaPart12008,Saw2014,BecSticky2013}. 
At the same time,
it is important to emphasize that the fundamental mechanism underlying these
processes---namely, the preferential concentration of inertial particles in
strain-dominated regions---remains qualitatively unchanged. What is modified,
however, are the quantitative measures of clustering (for example, the
correlation dimension $D_2$) and the associated statistics, which now depend on
the strength of the feedback. Thus, our results suggest a refinement, rather
than a breakdown, of the existing theoretical framework developed for dilute
suspensions.  A quantitative assessment of these effects would require direct
measurements of collision kernels, which are beyond the scope of the present
study.  Furthermore, the evidence of modifications in the small-scale structure
of the flow may lead to interesting effects on the nature of Lagrangian chaos,
irreversibility and related questions in turbulent
transport~\cite{BecLyapunov2006,BhatnagarPers2016,BhatnagarIrreversibility2018,RayIrreversibility2018}.}
Of course, the effect of such two-way couplings on the more \textit{real}
problem of particles in three-dimensional turbulence remains to be seen. This
is because, as we know, earlier studies have confirmed the dominant role of
intermittency in determining coalescence rates. Could the feedback term have a
negligible effect on this, or is it possible that the inclusion of this term
leads merely to an effective large Reynolds number flow? Nevertheless, it seems
reasonable to conjecture and explore in future studies the effect of such a
force on mixed-phase flows, such as those with filaments or polymer additives
in particulate suspensions. As we have seen, the particles do indeed enhance
small-scale straining (and vortical) regions, which in turn could lead to
enhanced stretching and eventual break-up~\cite{VincenziScission2021} of
polymers which coil and uncoil in such regions, or indeed change the dynamics
of long-chained filaments, which are sensitive to the local topology of the
flow field~\cite{PicardoChains2018,PicardoCollisions2019}.

\begin{figure}
	\includegraphics[width=0.98\linewidth]{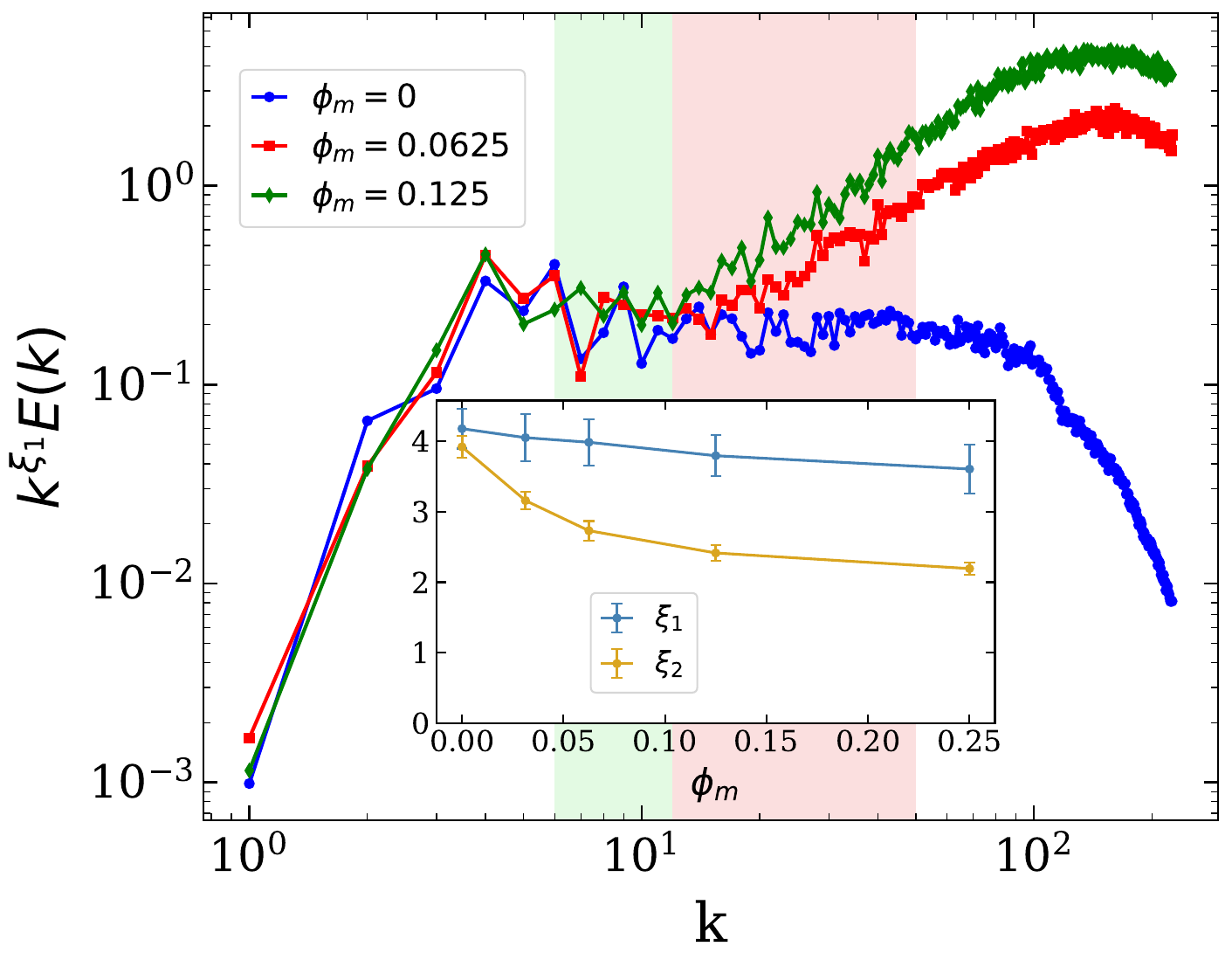}
\caption{Compensated energy spectra $k^{\xi_1}E(k)$ for different $\phi_m$. Two scaling regimes are observed: $E(k) \sim k^{-\xi_1}$ for $k_f \lesssim k \lesssim k_I$ and $E(k) \sim k^{-\xi_2}$ for $k_I \lesssim k \lesssim k_\eta$. 
\textcolor{black}{Inset: $\xi_1$ and $\xi_2$ as functions of $\phi_m$.}
}
\label{fig:Espectra}
\end{figure}

Second, the possible lack of uniqueness in the second-order inertial scaling
range $\zeta_2$ (Fig.~\ref{fig:intermittency}(a)) is already evident in the dual-scaling of the kinetic energy
spectrum $E(k) \sim k^{-\xi_1}$ for $k_f \lesssim k \lesssim k_I$ and $E(k) \sim
k^{-\xi_2}$ for $k_I \lesssim k \lesssim k_\eta$, as reported by Pandey
\textit{et al.}~\cite{PandeyDusty2019}; see also Refs.~\cite{SquiresDusty1990,BoivinDusty1998,Ferrante2003}; here $k_I$ corresponds to some intermediate cross-over
wavenumber and $k_\eta$ is associated with the small scales in the flow. 
We show in Fig.~\ref{fig:Espectra} a compensated kinetic energy spectrum
$k^{\xi_1} E(k)$, for different $\phi_m$, with two distinct scaling regimes.
In the first (green shaded region), the energy spectrum exhibits a more robust
scaling exponent $\xi_1 \approx 4$, which seems largely independent of $\phi_m$;
the second scaling regime, however, is characterized by $\xi_2$, which depends
quite strongly on $\phi_m$. In the inset of Fig.~\ref{fig:Espectra}, we plot
$\xi_1$ and $\xi_2$ as functions of $\phi_m$ to highlight this dependence
clearly. More importantly, a comparison of the scaling ranges for $\zeta_2$ and $\xi_2$ shows that the
dominant scaling exponent $\zeta_2$ stems from $\xi_2$; however, given the values of $\xi_2$, a trivial 
connection between these scaling exponents, through a Fourier transform, is hard to establish.

Nevertheless, the issue of multiple scaling ranges~\cite{PandeyDusty2019} is
vexing. What could be the possible origins of this behavior from the point of view of
standard turbulence phenomenology, and how could this be modeled in a
fluid-alone regime? We recall that for forced, statistically steady 2D
turbulence, there are two cascades: the inverse cascade of energy from the
forcing wavenumber $k_f$ all the way to the largest scales (smallest wavenumber
$k_L$), and a forward cascade of enstrophy from the forcing wavenumber $k_f$ up
to the largest wavenumbers $k_\eta$, beyond which enstrophy dissipation
dominates. These two cascades give rise to the well-known dual scaling regime:
$E(k) \sim k^{-5/3}$ for $k_L \lesssim k \lesssim k_f$ and $E(k) \sim
k^{-\xi_1}$ for $k_f \lesssim k \lesssim k_\eta$, with $3 \lesssim \xi_1
\lesssim 5$, a non-universal exponent which depends on the precise details of
forcing and dissipation~\cite{BoffettaReview2012,Nam2D2000,Boffetta2D2005,Tsang2D2005,Ray2D2011,Pandit2D2017}.

\begin{figure}
	\includegraphics[width=1.0\linewidth]{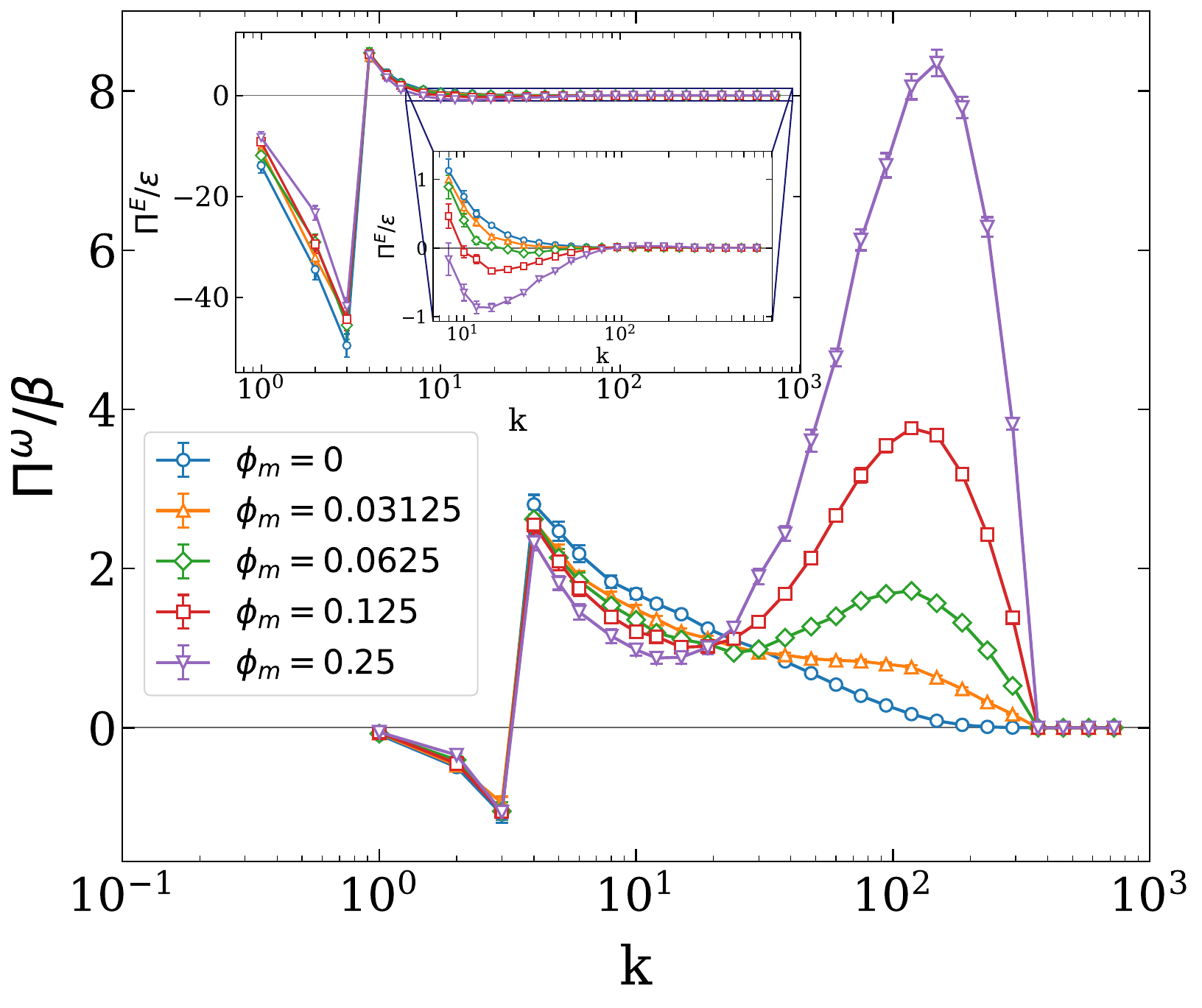}
\caption{Cumulative enstrophy flux (main panel) and energy flux (inset), normalized by the enstrophy ($\beta$) and energy ($\epsilon$) dissipation rates, respectively, for different $\phi_m$. 
\textcolor{black}{The inset shows a zoom for $k > k_f = 4$, highlighting an incipient inverse energy cascade that strengthens with increasing $\phi_m$.}
Error bars are obtained via bootstrapping.}
\label{fig:flux}
\end{figure}

For dusty turbulence, the scaling range due to the inverse cascade of energy is
inconspicuous because of the choice of the large-scale forcing at $k_f = 4$.
However, there is a second small-scale forcing $\boldsymbol{F}^d(\bx, t) \neq
0$ at large wavenumbers $k \sim \mathcal{O}(k_\eta)$ which can serve as a
source for inverse energy cascade at wavenumbers $k \lesssim k_f$ competing with
the direct enstrophy cascade from the external drive $f(\bx)$ at large scales.
Could this be the origin of the second scaling regime $\xi_2$ in the energy spectrum?

Given that the dust feedback acts as a distribution of point forces on the
fluid, this dual-scaling suggests that a dual-forcing model may offer a
reasonable approximation of the dust's momentum exchange with the carrier flow.
For such a model to be valid, however, the dusty flow must exhibit a cumulative
negative energy flux at high wavenumbers, indicative of an inverse cascade
driven by the small-scale (dust-induced) forcing.

We examine this effect from our DNS data by calculating the cumulative 
enstrophy flux 
\begin{equation}
    \label{eq:fluxEq}
\Pi^{\omega}(k) \equiv \langle \sum_{m\leq k} \widehat{\omega}_m \widehat{(\boldsymbol{u}\cdot \boldsymbol{\nabla}\omega)}_{-m} \rangle.
\end{equation}

In Fig.~\ref{fig:flux}, we show a plot of the cumulative enstrophy flux
$\Pi^{\omega}(k)$, normalised by the enstrophy dissipation rate $\beta$ for
various mass-loadings. There is indeed a sharp rise in the flux at large wavenumbers for $\phi_m \neq
0$, showing that dust particles enhance enstrophy transfer by injecting
small-scale vortical structures.

\textcolor{black}{Given this modification of the forward enstrophy cascade, a natural question is how strong the evidence is for a corresponding emergence of an inverse energy cascade. To address this,}
in the inset of Fig.~\ref{fig:flux}, we show a plot of the cumulative energy flux, normalised
by the energy dissipation rate $\epsilon$, for different $\phi_m$. A
cursory glance suggests that the effect of mass loading is not significant.
However, a zoomed-in view tells a slightly different story: there is a noticeable, if small, inverse energy cascade arising from the small-scale
forcing induced by the particles. \textcolor{black}{This suggests that particle-induced small-scale forcing enhances forward enstrophy transfer while simultaneously inducing a weak inverse transfer of energy, leading to the observed dual-scaling behavior in the energy spectra.} All of this opens up an intriguing
possibility of a two-scale forcing as an effective description of the fluid
phase of dusty turbulence, reminiscent of previous studies of multiscale forcing
in the two-dimensional Navier-Stokes equation~\cite{Mazzino2DPowerlaw2007}, and
the question of what happens in three-dimensional
turbulence~\cite{SainRandomForced1998} remains wide open. 
\textcolor{black}{However, the explicit representation of particle feedback as anisotropic,
point-like, and spatially correlated forcing remains intrinsically
complex due to the preferential concentration of particles. This
naturally motivates the use of an effective, coarse-grained forcing
framework, as we introduce below.}

\begin{figure*}
	\includegraphics[width=1\linewidth]{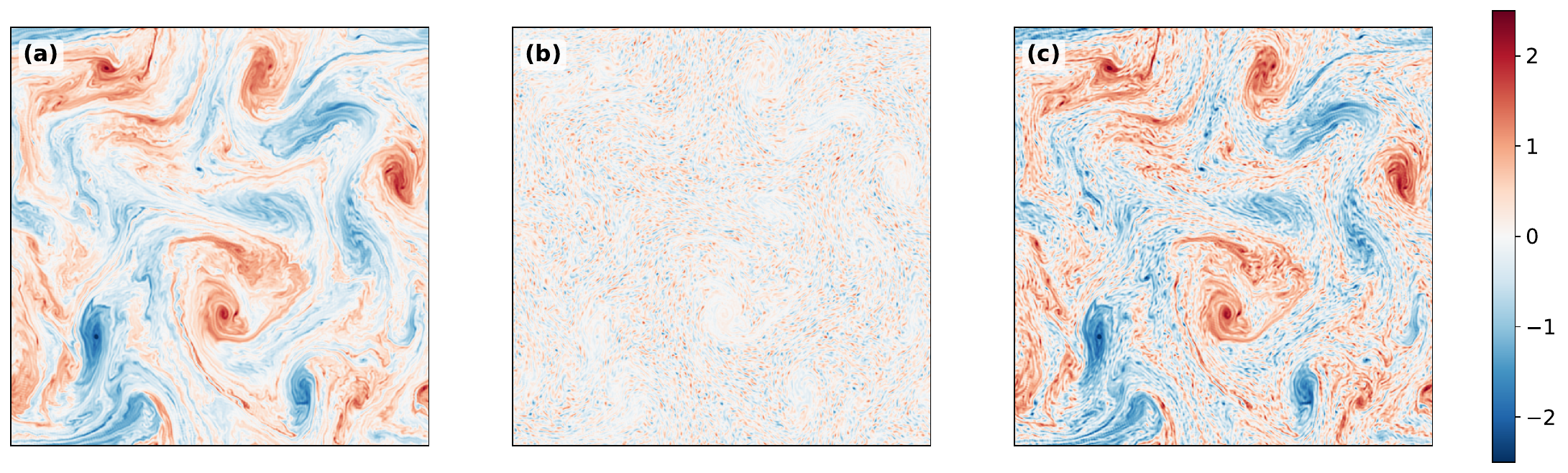}
\caption{\textcolor{black}{Representative vorticity fields from the masked dual-forcing model: (a) large-scale component $\omega_L$, (b) small-scale component $\omega_S$, and (c) total field $\omega = \omega_L + \omega_S$. 
The small-scale forcing generates localized, filamentary structures preferentially concentrated in strain-dominated regions, qualitatively similar to the small-scale features observed in particle-laden simulations (Fig.~2(b)).}
}
	\label{fig:vorticity_masking}
\end{figure*}

\begin{figure}
	\includegraphics[width=1\linewidth]{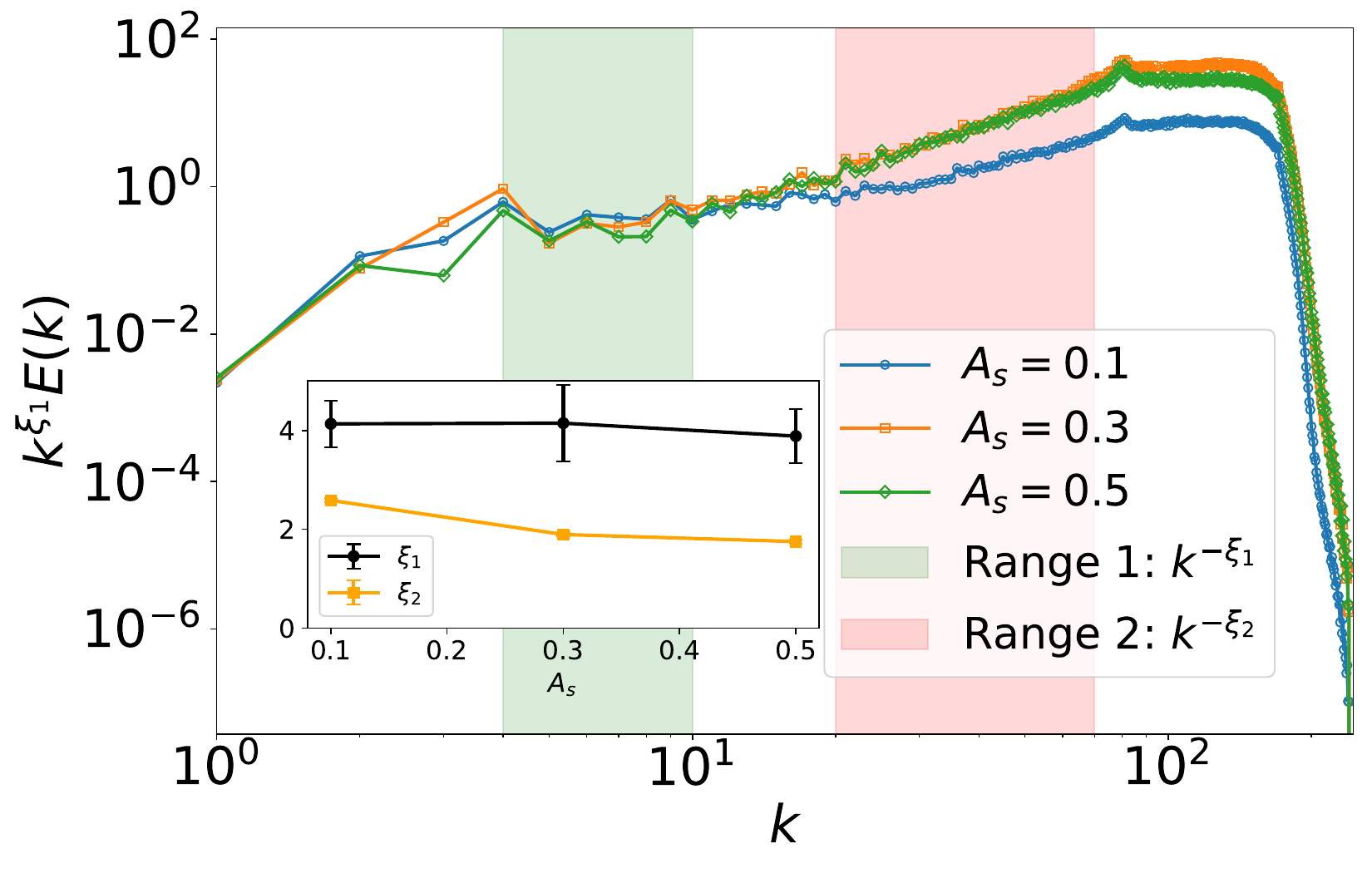}
\caption{\textcolor{black}{Compensated energy spectra for different small-scale forcing amplitudes $A_s$ at fixed masking parameter $\gamma = 10$. 
Two scaling regimes are evident, analogous to the dual-scaling behavior observed in the dusty turbulence simulations (Fig.~6). Increasing $A_s$ strengthens the secondary scaling regime, mirroring the effect of increasing mass loading $\phi_m$.}
}
	\label{fig:masked_dual_energy_spectrum}
\end{figure}

\begin{figure}
	\includegraphics[width=1\linewidth]{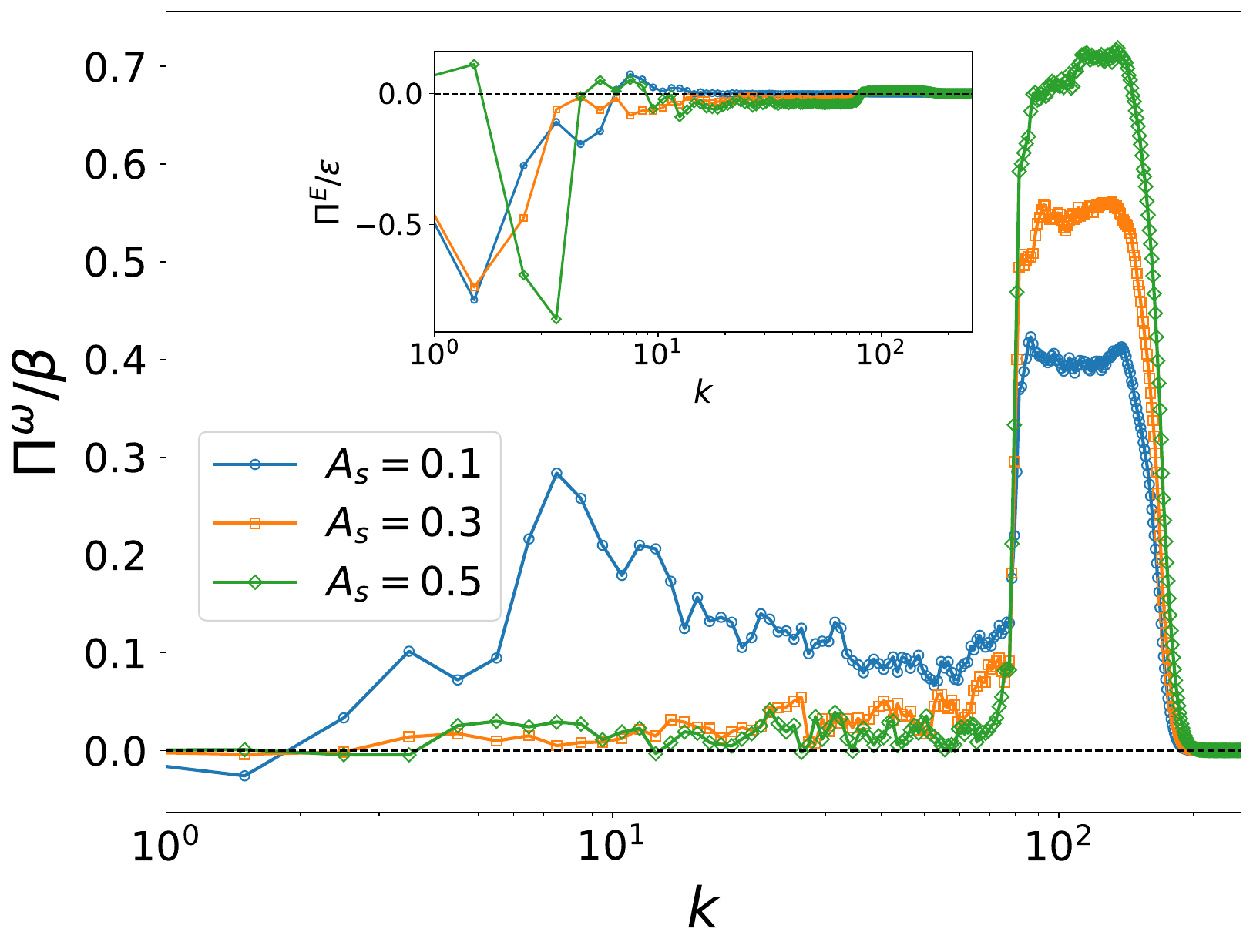}
\caption{\textcolor{black}{Cumulative enstrophy flux (main panel) and energy flux (inset) for different values of the small-scale forcing amplitude $A_s$ at fixed $\gamma = 10$. 
Increasing $A_s$ enhances the forward enstrophy transfer and induces a weak inverse energy flux at high wavenumbers, consistent with the trends observed in the particle-laden system (Fig.~7).}
}
	\label{fig:masked_dual_flux}
\end{figure}

\begin{figure}
	\includegraphics[width=1\linewidth]{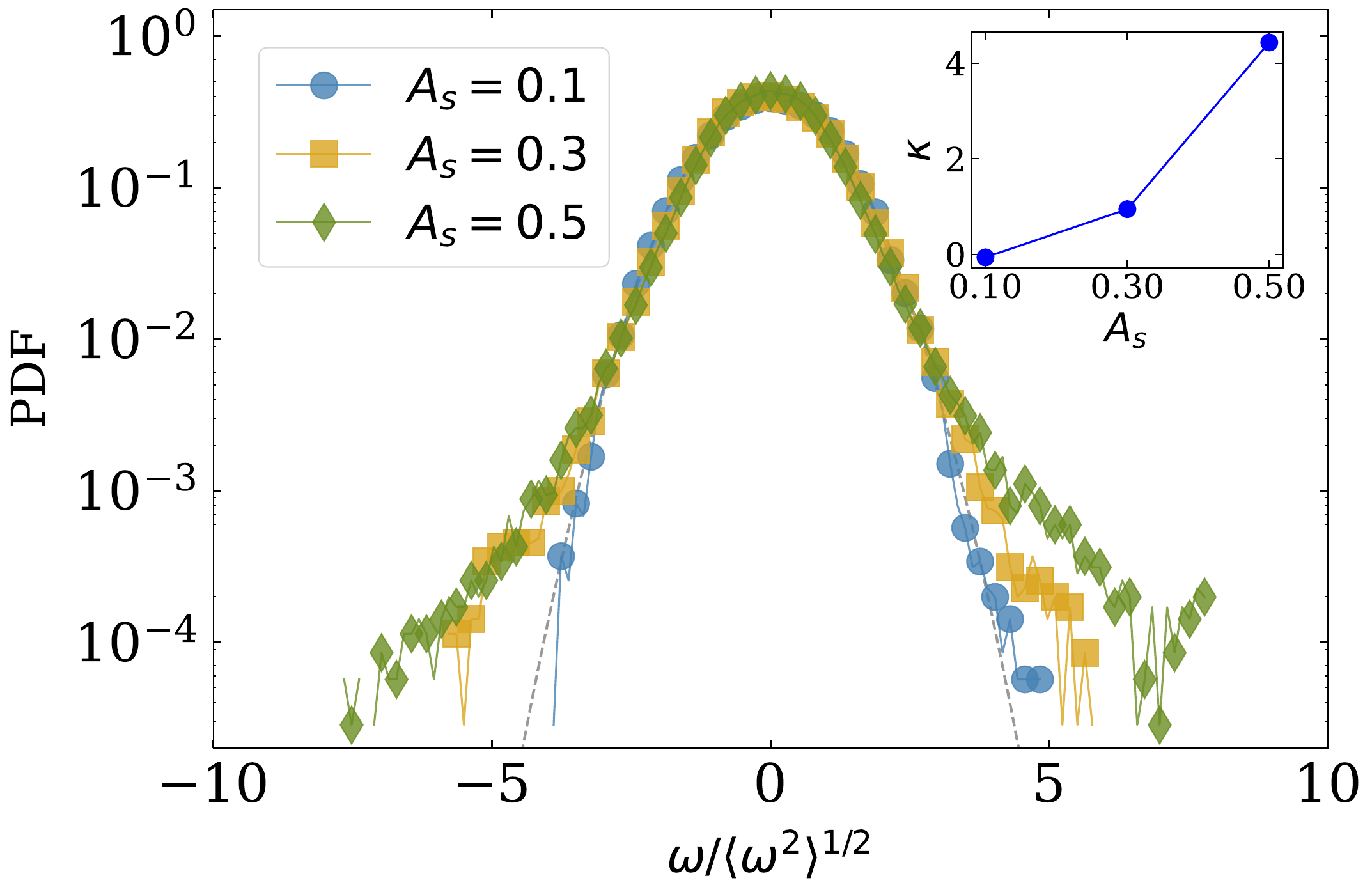}
\caption{\textcolor{black}{Probability density functions of the normalized vorticity for different small-scale forcing amplitudes $A_s$ at fixed masking parameter. 
Increasing $A_s$ leads to a systematic broadening of the distribution and (inset) an increase in kurtosis, indicating enhanced intermittency. This trend closely parallels the behavior observed as a function of mass loading $\phi_m$ (Fig.~2(c)).} }
	\label{fig:masked_pdf}
\end{figure}

{\color{black}
Following the observation of emergent dual-scaling regimes and an incipient inverse energy cascade driven by particle-induced small-scale forcing (Figs.~\ref{fig:Espectra} and \ref{fig:flux}), we propose a multiscale forcing model as an effective Eulerian description of the carrier fluid in dusty turbulence. The central idea is that particle back-reaction acts, at a coarse-grained level, as a source of small-scale forcing.

We consider the two-dimensional incompressible Navier-Stokes equations in vorticity form on a periodic domain:

\begin{equation}
	\partial_{t}\omega+\bu\cdot\nabla\omega=\nu(-\Delta)^{p}\omega-\alpha\omega+f_{L}+f_{S},
\end{equation}
with the velocity field given by
\begin{equation}
	\bu=\nabla^{\perp}\Delta^{-1}\omega, \quad \nabla\cdot \bu=0.
\end{equation}

Here, $f_{L}$ and $f_{S}$ denote statistically independent forcings acting at large and small scales, respectively. To achieve a well-resolved extended inertial range,equation is modeled using a hyperviscosity of order $p=8$ with $\nu=1$. An ekman friction term is added with $\alpha=0.1$ to avoid large-scale energy accumulation. In Fourier space, the forcing for components $i \in\{L,S\}$ is defined as

\begin{equation}
	\hat{f}_i(k,t) = 
	\begin{cases} 
		A_i \hat{\xi}_i(k,t), & k_i - \Delta k_i \le |k| \le k_i + \Delta k_i \\ 
		0, & \text{otherwise}
	\end{cases}
\end{equation} 
Here, $\hat{\xi}_{i}(k,t)$ are independent Gaussian, delta-correlated random variables satisfying $\langle\hat{\xi}_{i}(k,t)\hat{\xi}_{j}^{*}(k^{\prime},t^{\prime})\rangle\propto\delta_{ij}\delta_{kk^{\prime}}\delta(t-t^{\prime})$. The large-scale forcing, driving the primary flow dynamics, is concentrated within a low-wavenumber shell centered at $k_L=3$ with $\Delta k_L=2$. To model the back-reaction of inertial particles at a coarse-grained level, the small-scale forcing is injected at a high wavenumber $k_S=80$ with $\Delta k_S=2$. The scale-specific amplitudes $A_{i}$ are dynamically chosen to impose constant mean energy injection rates, $\epsilon_{i}=\langle u_{i}\cdot f_{i}\rangle$.

To isolate the dynamical effects of the distinct forcing scales, we decompose the total vorticity and velocity fields linearly as
\begin{equation}
	\omega=\omega_{L}+\omega_{S}, \quad \bu=\bu_{L}+\bu_{S}.
\end{equation}
The evolution of these component fields are governed by
\begin{equation}
	\partial_{t}\omega_{L}+\bu\cdot\nabla\omega_{L}=\nu(-\Delta)^{p}\omega_{L}-\alpha\omega_{L}+f_{L},
\end{equation}
\begin{equation}
	\partial_{t}\omega_{S}+\bu\cdot\nabla\omega_{S}=\nu(-\Delta)^{p}\omega_{S}-\alpha\omega_{S}+f_{S}.
\end{equation}
The respective velocity fields are obtained independently via
\begin{equation}
	\bu_{i}=\nabla^{\perp}\Delta^{-1}\omega_{i}, \quad i\in\{L,S\}.
\end{equation}

In the particle-laden system, however, the feedback forcing is neither
spatially homogeneous nor isotropic: inertial particles preferentially sample
the flow and cluster in strain-dominated regions, leading to a localized and
topology-dependent forcing on the carrier fluid. To more accurately mimic this
anisotropic, localized feedback, we introduce a spatial masking function to the
small-scale forcing.

The modified coupled system is written as
\begin{equation}
	\partial_{t}\omega_{S}+\bu\cdot\nabla\omega_{S}=\nu(-\Delta)^{p}\omega_{S}-\alpha\omega_{S}+f_{S} \mathcal{M}(x,y),
\end{equation}
where the masking function $\mathcal{M}(x,y)$ restricts the application of $f_{S}$ based on the local flow topology. We define this mask using the Okubo-Weiss parameter of the large-scale flow, $\Lambda_L$, which distinguishes between vorticity-dominated and strain-dominated regions:
\begin{equation}
	\Lambda_L = \omega_L^2 - \sigma_{n,L}^2 - \sigma_{s,L}^2,
\end{equation}
where $\omega= \partial_x u_{y} - \partial_y u_{x}$, $\sigma_{n} = \partial_x u_{x} - \partial_y u_{y}$ and $\sigma_{s} = \partial_x u_{y} + \partial_y u_{x}$ denote the normal and shear strains, respectively and the subscript $L$ denote the large-scale flow. The mask is then constructed as
\begin{equation}
	\mathcal{M}(x,y) =\frac{1}{2}\left[1- \tanh\left(\gamma \frac{\Lambda_L}{\Lambda^{rms}_L}\right)\right],
\end{equation}
where $\Lambda^{rms}_L$ is the root-mean-square of $\Lambda_L$, and $\gamma$ is a sharpness parameter that tunes the spatial extent of the regions where the small-scale forcing $f_S$ is active.

This construction allows us to interpolate between uniform and highly localized
small-scale forcing, thereby mimicking the preferential concentration of
inertial particles. A representative decomposition of the resulting flow into
large-scale, small-scale, and total vorticity fields is shown in
Fig.~\ref{fig:vorticity_masking}, where the small-scale component exhibits
localized, filamentary structures qualitatively similar to those observed in
the particle-resolved simulations (Fig.~\ref{fig:Eu1ptVort}(b)).

We now examine the statistical consequences of this effective forcing. In
Fig.~\ref{fig:masked_dual_energy_spectrum}, we show the compensated energy
spectra for different amplitudes of small-scale forcing $A_s$. As $A_s$
increases, a secondary scaling regime emerges at high wavenumbers, in close
analogy with the dual-scaling behavior observed in Fig.~\ref{fig:Espectra}, as the mass loading
$\phi_m$ is increased.

Further insight is obtained from the spectral fluxes shown in
Fig.~\ref{fig:masked_dual_flux}. Increasing $A_s$ enhances the forward
enstrophy transfer and induces a weak inverse energy flux at high wavenumbers,
consistent with the trends observed in the particle-laden system (Fig.~\ref{fig:flux}). This
supports the interpretation that particle feedback effectively injects energy
at small scales.

Finally, in Fig.~\ref{fig:masked_pdf}, we show the PDFs of the normalized
vorticity field for different values of $A_s$. Increasing $A_s$ leads to a
systematic broadening of the distributions and an increase in intermittency levels,
closely paralleling the behavior observed as a function of $\phi_m$ in
Fig.~\ref{fig:Eu1ptVort}(c). While the extreme tails remain sensitive to resolution and
statistical convergence, the trend of increasing intermittency is robust.

Taken together, these results indicate that the dominant effect of particle
feedback on the carrier flow can be understood, at leading order, as an
effective, spatially localized small-scale forcing. Although a fully resolved
representation of particle-induced forcing remains challenging due to its
anisotropic and point-like nature, the present model provides a minimal
coarse-grained framework that captures the key statistical signatures of dusty
turbulence. Ultimately, this dual-scale forcing mechanism serves as a
computationally efficient Eulerian description of particle-laden turbulence,
which may be extended to more complex multiphase flows.}

We thank Prasad Perlekar for useful discussions.
S.S.R. acknowledges SERB-DST (India) projects the CEFIPRA Project No 6704-1 for support.  This research was supported in part by the International Centre for
Theoretical Sciences (ICTS) for the program 10th Indian Statistical Physics
Community Meeting (code: ICTS/10thISPCM2025/04).  The simulations were
performed on the ICTS clusters Mario, Tetris, and Contra.  The authors
acknowledge the support of the DAE, Government of India, under projects nos.
12-R\&D-TFR-5.10-1100 and RTI4001.

\bibliographystyle{apsrev4-2} 
\bibliography{references}

\end{document}